\documentclass[debug,overfull]{epl}
\usepackage{amsmath,graphicx}
\title{Dynamics of a polymer in a quenched random medium: A Monte Carlo investigation}
\author{A.
  Milchev\inst{1,2}\thanks{E-mail: \email{milchev@mail.uni-mainz.de}}
  \and V.G. Rostiashvili \inst{1}\thanks{E-mail: \email{rostiash@mpip-mainz.mpg.de}} \and
  T.A. Vilgis \inst{1}\thanks{E-mail: \email{vilgis@mpip-mainz.mpg.de}}}
\institute{
  \inst{1} Max - Planck - Institute for Polymer Research -
  Ackermannweg 10, 55128 Mainz, Germany\\
  \inst{2} Institute for Physical Chemistry, Bulgarian Academy of
  Sciences, 1113 Sofia, Bulgaria 
}

\begin{document}

\maketitle

\begin{abstract}
We use an off - lattice bead - spring model of a self - avoiding 
polymer chain immersed in a $3$-dimensional quenched random medium 
to study chain dynamics by means of a Monte - Carlo (MC) simulation.

The chain center of mass mean-squared displacement as a function of time
reveals two crossovers which depend both on chain length $N$ and on the degree
of Gaussian disorder $\Delta$. The first one from normal to anomalous
diffusion regime is found at short time $\tau_1$ and observed to vanish
rapidly as $\tau_1 \propto \Delta^{- 11}$ with growing disorder.  The second
crossover back to normal diffusion, $\tau_2$, scales as $\tau_2 \propto
N^{2\nu + 1} f(N^{2-3\nu}\Delta)$ with $f$ being some scaling function.  The
diffusion coefficient $D_N$ depends strongly on disorder and drops
dramatically at a {\em critical dispersion} $\Delta_{c} \propto N^{-2 + 3\nu}$
of the disorder potential so that for $\Delta > \Delta_c$ the chain center of
mass is practically frozen.

The time-dependent Rouse modes correlation function $C_{p}(t)$ reveals a
characteristic plateau at $\Delta > \Delta_c$ which is the hallmark of a non -
ergodic regime. These findings agree well with our recent theoretical
predictions.
\end{abstract}

\section{Introduction}
Considerable progress has been made in understanding of the static
\cite{Edward,Cates,Machta,Machta1,Natter,Obukh,Doussal,Stepan,Vilgis,Gold,Gold1}
and dynamic \cite{Stepan1,Ebert,Ebert1} behavior of polymer chains in a
quenched random medium. It was shown already in the paper by Cates and Ball
\cite{Cates} that for sufficiently large diffusion coefficient (so as the
chain may sample the whole system) the quenched and annealed problems are
equivalent.  Indeed, since the chain always moves towards the deepest energy
minimum, quenched and annealed problems can differ only for finite volume of
the random medium\cite{Gold,Gold1}.

In a recent publication \cite{Miglior} we have argued that dynamic aspects can
significantly affect the problem in question.  We have treated this
problem\cite{Miglior} using the generating functional method which based on
the Langevin dynamics. Within this method the averaging over the quenched
disorder leads to coupling between different dynamic trajectories which in
turn eventually results in a non - Markovian diffusional slowing down, i.e.in
anomalous diffusion, as well as in non - ergodic behavior for the Rouse modes
of the polymer chain. The resulting equation of motion (which was derived
within the Hartree approximation) has a memory kernel typical for the mode -
coupling theory of the glass transition \cite{Gotze}. Self - consistent
treatment of this equation leads to the conclusion that in $d$ - dimensional
space the center of mass diffusion coefficient decreases according to the law:
\begin{equation}
D_N \approx D_{\rm R}[1 - {\rm const}(\Delta/b^d)N^{\alpha}]\quad , 
\label{Diffusion}
\end{equation}
where $\Delta$ is the dispersion of the disorder potential, $b$ is the Kuhn
segment length, and $\alpha = 2 - \nu d$ ($\nu$ is the Flory exponent). This
differs remarkably from the earlier prediction, $D_N \approx D_{\rm R} \exp[-
(\Delta/b^d)N^{\alpha}]$, based on a simple Markovian diffusion argumentation
\cite{Machta,Machta1,Miglior}.  As a result some critical disorder,
$\Delta_{\rm c} \approx b^d N^{- \alpha}$, exist such that at $\Delta >
\Delta_{\rm c}$ the chain center of mass is arrested by the random potential.
In the case when the capture time $t_{\rm cap} >> t_{\rm eq}$ (where $t_{\rm
  eq}$ is the time of internal degrees of freedom equilibration) the quenched
and annealed problems are not equivalent even for an infinite volume! Rouse
modes of the chain also freeze at a common disorder strength $\Delta$, so that
the non - ergodicity plateau shows up according the A - type (or continuous)
transition \cite{Gotze}. In the present letter we will mainly check these
important predictions of the ref.\cite{Miglior} by the direct and vast MC -
simulations. In the conclusion we also sketch some possible future
developments.

\section{Model} 
In the present study we employ an off-lattice Monte Carlo model of 
a bead-spring polymer chain,  used in numerous investigations before\cite{BA}.
During the simulation beads are chosen at random and displaced to a new 
trial position in their environment subject to a Metropolis procedure.
Each polymer chain consists of $N$ beads, connected
by anharmonic springs which are described by the finitely-extensible
nonlinear elastic (FENE) potential\cite{AMKB}

\begin{equation}
U_{FENE}=-\frac{KR^{2}}{2}\ln \left[ 1-\frac{\left( l-l_{0}\right) ^{2}}{%
R^{2}}\right] ,  \label{FENE}
\end{equation}
where $l$ is the bond length which can vary in between $l_{\min }<l<l_{\max
},$ and has the equilibrium value $l_{0},$ with $R=l_{\max }-l,\;l_{\min
}=2l-l_{\max },$ and $K$ is the spring constant. As before\cite%
{AMKB}, we take $l_{\max }=1$ as the unit length, $l_{\min
}=0.4,\;l_{0}=0.7,$ and $K/k_{B}T=40.$ The non-bonded interactions between
the beads are described by the Morse potential\cite{BA}

\begin{equation}
U_{M}(r)=\varepsilon _{M}\exp \left[ -2a \left( r-r_{\min }\right) %
\right] -\exp \left[ -a \left( r-r_{\min }\right) \right] ,
\label{Morse}
\end{equation}
where $r$ is the distance between the beads, $r_{\min }=0.8,\;\epsilon
_{M}=1,$ and $a =24$ (since the potential has a very short range, the
simulation code is rather fast\cite{AMKB}). The $\Theta $%
-temperature for the model is $k_{B}T_{\theta }\approx 0.62,$ so that at
$k_{B}T = 1$, used throughout in this study, the chain is in a good solvent
regime and all energetic variables are dimensionless.

The energetic disorder is created by dividing the simulation box into $64^3$
hypothetical subcells, each of size $l_{max}=1$, which are ascribed a
Gaussian-distributed local potential so that its dispersion is taken as a
measure of the degree of disorder $\Delta$.  Typically runs of length $10^8
\div 10^9$ MC steps (MCS) have been carried out and observables have been
averaged over $500$ different realizations of disorder.

\section{Results} First of all we emphasize that the stochastic
dynamics in the system under study is {\em not} time translational invariant 
(TTI) \cite{Bouchaud}. This means  that in the time - dependent
correlation function one should keep control on two times: the waiting 
time $t_{\rm w}$, i.e. the time elapsed after the system preparation
(aging) before any measurement starts, and the actual measurement
time $t$. With this in mind we have studied the center of mass mean -
square displacement (MSQD)
\begin{equation}
\label{displacement}
g_3(t,t_{\rm w}) = \langle [{\bf r_{CM}}(t) - 
{\bf r_{CM}}(t_{\rm w})]^2\rangle \quad,
\end{equation}
where the center of mass position ${\bf r_{CM}}(t) =
\sum_{i=1}^{N} {\bf R}_i(t)/N$ and the angular brackets
denote an average over the realizations of disorder. In contrast, in
cases of vanishing disorder TTI holds and the time correlation functions
depend on a single variable $\delta t = t-t_w$.

\begin{figure}[ht]
\onefigure[width=6cm,angle=270]{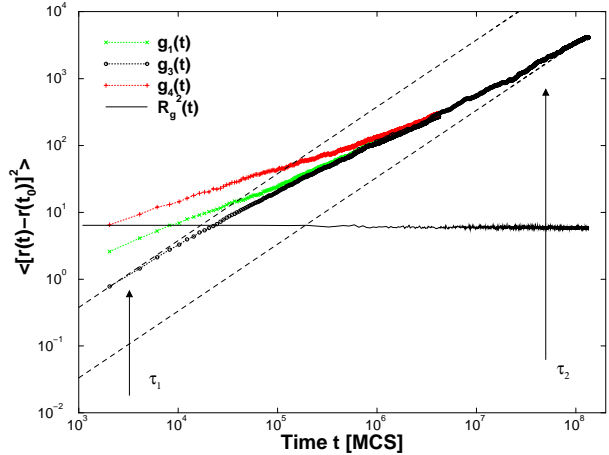}
\caption{Mean - square displacement of a central monomer, $g_1(t)$,
of an end monomer, $g_4(t)$, and of the chain center of mass, $g_3(t)$ vs. time
for $N=32$ and $\Delta=0.95$.
Dashed lines with slope $1$ denote normal diffusive behavior whereas
the crossover times $\tau_1$ (from) , and $\tau_2$ (back to) normal diffusion
are marked by arrows. The mean squared radius of gyration of the polymer, 
$R_{\rm g}^2(t)$, appears as a horizontal line.}
\label{MSD}
\end{figure}

Fig.\ref{MSD} shows a typical example for the chain length $N = 32$ the
disorder strength $\Delta = 0.95$ and the MC - time interval $t > 10^8$. One
can clearly distinguish three diffusional regimes.  In the initial short
interval $0 < t < \tau_1$ a relatively fast normal diffusion shows up. The
crossover time $\tau_1$ corresponds to the moment when disorder starts to
effect the chain motion. In ref.\cite{Miglior} this moment was denoted by
$\tau_{\rm d}$ and was defined as the point where the memory - friction term
overwhelms the bare - friction term. After that the non - Markovian diffusion
gets under way. In the interval $\tau_1 < t < \tau_2$ the MSQD manifests
itself as anomalous diffusion with $g_3(t) \propto t^{\beta}$, where the
exponent $\beta < 1$. At time $t > \tau_2$ (which corresponds to the MSQD
$g_3(t) \gg R_{\rm g}^2$)  the diffusion goes back to normal albeit with a
significant slowing down (see below).

\begin{figure}
\twofigures[scale=0.33,angle=270]{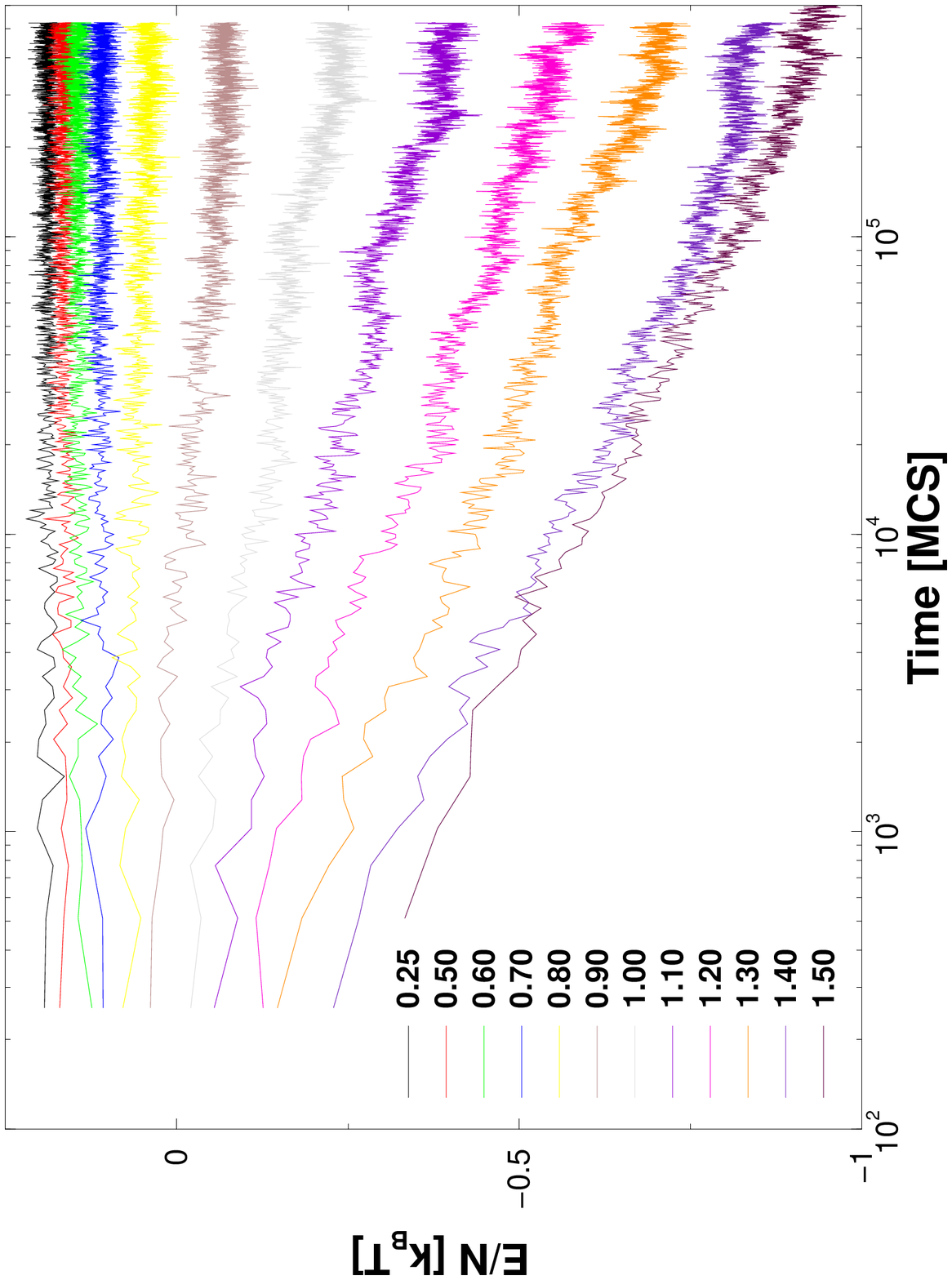}{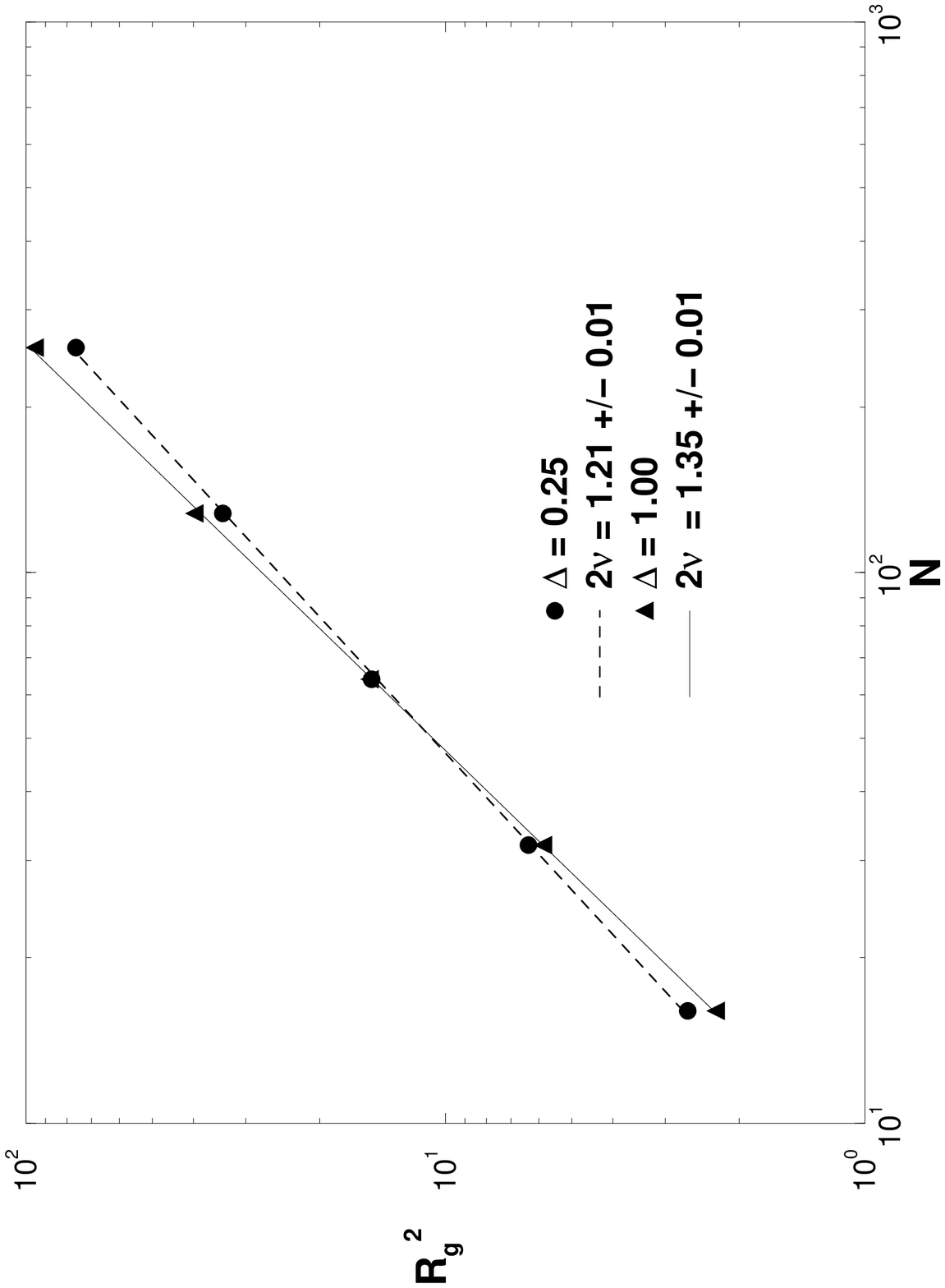}
\caption{Energy (per monomer) vs time of a chain with $N=8$ for growing
degree of disorder $\Delta$ (given as a parameter).}
\label{Drift}
\caption{A scaling plot of the radius of gyration, $R_g^2$, with chain
length $N$ for small, $\Delta=0.25$, and large, $\Delta=1.00$, degree of
disorder. }
\label{Radius}
\end{figure}

The MSQD - plot given in Fig.\ref{MSD} has been obtained for waiting 
time $t_{\rm w} = 0$. In the present report we shall not consider
aging phenomena (i.e. the $t_{\rm w}$ - dependence of the time correlation 
functions as well as the violation of the fluctuation - dissipation
theorem (FDT)) putting this off for a later publication. Nevertheless
Fig.\ref{Drift} clearly demonstrates the drift of the full energy $E$
with time $t$ which for $N=8$ gets pronounced at $\Delta > 0.9$.
Evidently in this case the TTI is violated for larger disorder. The age of the
system also influences static conformational properties as the mean - squared
radius of gyration $R_{\rm g}^2$ of the polymer chain.  Figure \ref{Radius}
shows that at small disorder, $\Delta = 0.25$, the $R_{\rm g}^2$ versus $N$
dependence has the same slope as in the pure (i.e. without any disorder)
system. An inspection of the measured $R_{\rm g}^2$ vs. $N$ relationship
yields $\nu = 0.605 \pm 0.005$ for the Flory exponent. At large disorder,
$\Delta = 1.00$, however, the slope of $R_{\rm g}^2(N)$ is affected by the
waiting time $t_{\rm w}$. The value of $t_{\rm w}$ which we have used in
Fig.\ref{Radius} corresponds to the slow diffusion regime, i.e. $t_{\rm w} >
\tau_2$. In this case $\nu = 0.68 \pm 0.01$, i.e. $\nu > \nu_{\Delta=0}$ which
is consistent with the results obtained on hierarchical lattices
\cite{Doussal} for the quenched problem. The quenched problem in
ref.\cite{Doussal} corresponds to the case in which the one end of the chain
is kept fixed. In this case the polymer stretches away from the fixed end in
order to find a region with a minimal energy. The Imry - Ma arguments have
been used in ref \cite{Obukh} to show that $\nu_{\rm quenched} = 2/3$.

\begin{figure}[ht]
\onefigure[width=6cm,angle=270]{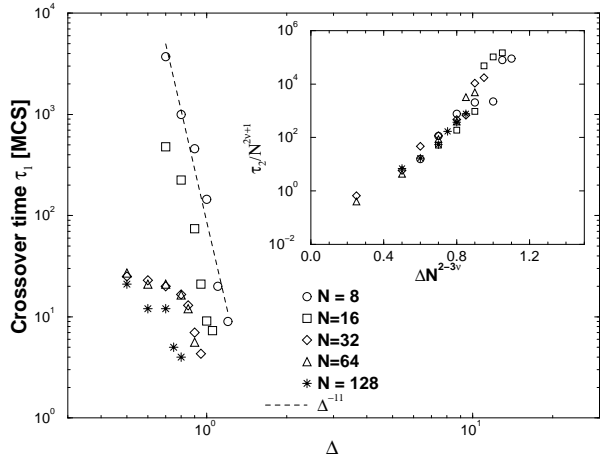}
\caption{Crossover times $\tau_1$ and $\tau_2$ (inset)  for chains 
of length $N$ at different degree of disorder $\Delta$.}
\label{Crossover}
\end{figure}

Now we go back to MSQD - plot (Fig.\ref{MSD}) to discuss how the
crossover times, $\tau_1$ and  $\tau_2$, depend on $\Delta$. This is
presented in Fig.\ref{Crossover}

The first crossover $\tau_1$ decreases whereas the second one $\tau_2$ 
increases with the disorder strength $\Delta$, so that the anomalous
diffusion interval is extended. While for the relatively
long chains, $N \ge 32$, the initial interval $\tau_1$ is getting
progressively so short that a scaling dependence on $\Delta$ is difficult to
establish, for shorter chains (where the statistics is better too) 
the $\tau_1$ vs $\Delta$ relationship can be reasonably fitted by 
$\tau_1 \propto \Delta^{- 11}$. In ref. \cite{Miglior} a theoretical 
prediction for $\tau_1$ was derived as:
\begin{equation}
\tau_1 \propto \Delta^{- 1/(1 - \beta)} \quad,
\label{tau_1}
\end{equation}
where the anomalous diffusion exponent $\beta = 1 - (2 - \nu d)/(2\nu
+ 1) < 1$. In the case of $d = 3$ and $\nu = 0.6$ one obtains $\beta = 0.9$
which leads to $\tau_1 \sim \Delta^{- 10}$. This prediction agrees
closely with the above - mentioned MC findings. As is clearly seen in the 
inset of Fig. \ref{Crossover}, all data for the second crossover time 
$\tau_2$ follow a scaling law $\tau_2 \propto N^{1+2\nu}f(\Delta 
N^{2-3\nu})$ where $f$ is some scaling function.

\begin{figure}[ht]
\onefigure[width=6cm,angle=270]{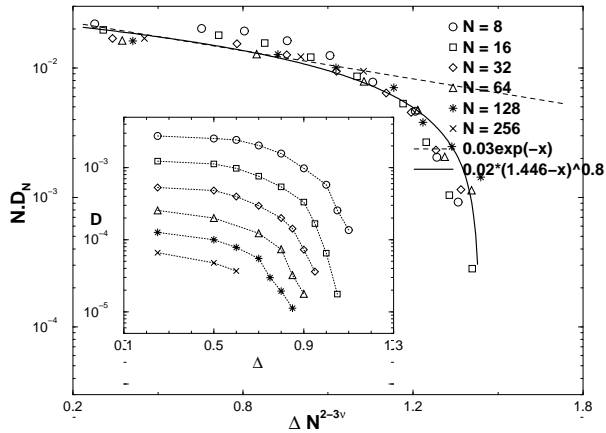}
\caption{Variation of the normalized diffusion coefficient $D_N/D_R$ with
scaled degree of disorder $x=\Delta N^{2-3\nu}$ for chains of length
$8\le N\le 256$. Dashed line denotes the simple Markovian diffusion
prediction  $ND_N
\propto \exp(-x)$\cite{Machta1,Miglior}, a full line is a best fit 
with $ND_N \propto (x_c - x)^{0.8}$. Unscaled data is shown in the inset.}
\label{Slow}
\end{figure}

As mentioned above, at $t > \tau_2$ the normal diffusion
regime shows a significant slowing down. This extremely important  result is
depicted in Fig.\ref{Slow} over a wide range of chain lengths.

All points in the reduced coordinates ($N D_N$ vs $\Delta N^{2 - 3 \nu}$) lie
on a single master curve.  It appears that our MC - data nicely follows the
best fit $D_N \propto N^{-1} (x_c - x)^{0.8}$ (where $x = \Delta N^{2-3\nu}$)
which correlates well with the theoretical predictions for $D_N$ (see
eq.(\ref{Diffusion})). A remarkable and strong systematic difference between
the master curve (solid line) and the simple Markovian diffusion result
(dashed line) can be seen, which grows progressively with increasing of degree of
disorder.  The basis for this slowing down is the coupling between diffusional
trajectories (or between different parts of the same trajectory) caused by the
averaging over the disorder. This furnishes the non - Markovian slowing down
regime, which mathematically shows up as a memory kernel term in the resulting
equation of motion \cite{Miglior}. Such behavior is generic for the
``schematic'' mode - coupling theory of the glass transition \cite{Gotze}. The
existence of such dynamic singularity for a finite chain is in principle
questionable. More precisely one should better mean that the capture time of
the center of mass, $t_{\rm cap}$, is finite albeit astronomical, so that in
the slowing down regime $t_{\rm cap} \gg t_{R} \exp(\Delta N^{2-3\nu})$.  This
time scale goes beyond the scope of our present MC - simulation.

\begin{figure}[ht]
\onefigure[width=6cm,angle=270]{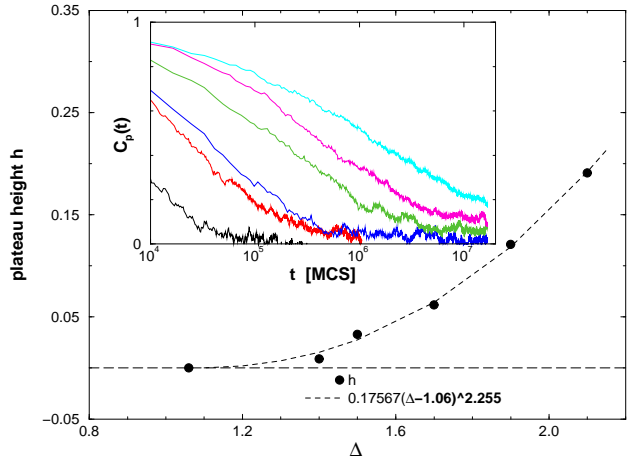}
\caption{Variation of the plateau height $h$ with degree of disorder 
$\Delta$ of the Rouse-mode correlation function $C_1(t,0)$ of a chain with
$N=16$. In the inset
$C_1(t,0)$ are shown for the respective values of $\Delta$.}
\label{Rouse_correlator}
\end{figure}

So far, we have focused mainly on the chain center of mass slow dynamics.  We
studied also the time dependent correlation function of the Rouse modes
\cite{Doi}
\begin{equation}
\label{Rouse}
C_{\rm p}(t, t_{\rm w}) = \left< X_{\rm p}(t) X_{\rm p}(
  t_{\rm w})\right> \quad,
\end{equation} 
where the Rouse mode component with an index $p$, $X_p(t)=N^{-1}
\sum_{i=1}^N{\bf r}_i(t)\cos[(i-0,5)p\pi N^{-1}]$, is measured at time $t$. This correlation function reflects
the dynamics of the internal degrees of freedom. It can be seen in Fig.
\ref{Rouse_correlator} that at relatively large disorder strength $\Delta$ a
characteristic plateau appears which is reminiscent of a non - ergodic
behavior for the so called $A$ - type dynamical transition \cite{Gotze}.
Figure \ref{Rouse_correlator} illustrates this non - ergodic plateau behavior
for the first mode ($p = 1$) and the chain length $N = 16$.

The hight $h$ of the plateau is found to be a continuous function of $\Delta$,
i.e. $h \propto (\Delta - 1.06)^{2.5}$, which qualitatively agrees
with the recent theoretical predictions \cite{Miglior}. It is 
interesting to note that the critical $\Delta_{\rm c} \approx 1.06$ coincides
with the critical value $\Delta_{\rm c}$ which is
obtained for $N = 16$ from the diffusion results (see
the Fig.\ref{Slow}). The well - known $F_1$ model, which has been
studied in ref.\cite{Gotze}, provides  a simple example of  such kind of
continuous transition.

\section{Conclusion} In this work an off - lattice model of a 
self - avoiding polymer chain in quenched random field is used 
to study the slow  polymer dynamics. We examine the center of mass 
MSQD at different degree of disorder and find a
broad regime of anomalous diffusion which starts at a crossover
time $\tau_1$ and ends at a second crossover $\tau_2$ by going into 
normal diffusion regime. We have demonstrated that these crossover 
times are observed to scale in agreement with the theoretical predictions
\cite{Miglior}. 

The diffusion coefficient $D_N (\Delta)$ at $t > \tau_2$ drops dramatically at
a critical value $\Delta_{\rm c} \propto N^{-2 + 3 \nu}$. Simulation data for
$D_N$ of different polymer chain length and different degree of disorder is
observed to collapse on a single master curve as predicted in a recent theory
\cite{Miglior}. This significant slowing down effectively originates from the
coupling between different dynamic trajectories of a polymer chain caused by
the averaging over many disorder realizations. The Hartree approximation which
has been used in \cite{Miglior} is close in spirit to the well- known mode -
coupling approximation \cite{Gotze} and provides the basis for the treatment
of non - Markovian regimes. In contrast, a treatment in terms of Markovian
arguments yields an exponential relationship between diffusion coefficient and
reduced disorder which proves to be inadequate. The same reasons bring about
the Rouse modes freezing at a common disorder strength $\Delta_{\rm c}$,
whereby the generic non - ergodicity plateau appears continuously. This is
also in line with the theoretical predictions \cite{Miglior}.

Finally we should stress that the system at issue does not satisfy TTI so that
the waiting time $t_{\rm w}$ is a very important parameter. Although $t_{\rm
  w}$ has been controlled in the calculation of MSQD (eq.(\ref{displacement}))
and the Rouse mode correlators, eq.(\ref{Rouse}), its impact on the dynamics
is to be reported in a following communication. To this end the generating
functional approach, which has been used in ref.\cite{Miglior}, should be
generalized to the out - of - equilibrium dynamics \cite{Bouchaud} where the
influence of $t_{\rm w}$ (or the aging) and FDT violation are the effects of a
primary interest. We plan to consider these problems in a future publication.

\acknowledgments
One of us, A. M. acknowledges the support and hospitality of the Max-Planck 
Institute for Polymer Research in Mainz during this study. This research has 
been supported by the Sonderforschungsbereich (SFB 625). The authors have 
benefitted from discussions with K. Kremer and S. Obukhov. Assistance in the 
initial stages of this work by G. Migliorini is gratefully acknowledged.

\end{document}